\documentclass[secnumarabic, amssymb, amsmath,amsfonts,nobibnotes, aip, jmp]{revtex4}

\begin{document}

\title { Hadamard Matrices from Mutually Unbiased Bases }

\author{P. Di\c t\u a }
\email{dita@zeus.theory.nipne.ro}
\affiliation{National Institute of Physics and Nuclear Engineering,\\
P.O. Box MG6, Bucharest, Romania}
\date{February 24, 2010}
\begin{abstract}

An analytical method for getting  new complex Hadamard matrices by using mutually unbiased bases and a nonlinear doubling formula is provided. The method is illustrated with the $n=4$ case that   leads to a rich family of eight-dimensional  Hadamard matrices that depend on five arbitrary phases.

\end{abstract}

\maketitle

\section{Introduction}

 Mutually unbiased bases (MUBs) constitute a basic concept of quantum information and play an essential role in quantum tomography, quantum cryptography, construction of dense coding schemes, teleportation, classical signal processing, etc. Its origin is in the Schwinger paper \cite{Sc}.

Two orthonormal bases in $\mathbb{C}^d$, $A=(a_1,\dots,a_d)$ and $B=(b_1,\dots,b_d)$, are called MUBs if the moduli of the scalar product of theirs vectors $a_i$ and $b_j$ satisfy the relations $|\langle a_i,b_j\rangle|=\frac{1}{\sqrt{d}}, \;1\le i,j\le d$. An equivalent definition is that the product $A\,B^*$of the two complex Hadamard matrices generated by  $A$ and $B$ is again a Hadamard matrix, where $*$ denotes the Hermitian conjugate.

It is well known that MUBs exist for any dimension $d$, a known result being that for any $d\ge 2$ there exist at least three MUBs, see \cite{Iv}, \cite{Po},\cite{Co}. It was also established that the maximum cardinality of a set of MUBs in $\mathbb{C}^d$ is less or equal to $d+1$, with equality for $d=p^n$, a power of a prime number \cite{WF}, \cite{Iv}, \cite{BBRV}, \cite{KR}. For non prime numbers the results are very scarce such that even the simplest   case, $d = 6$, is not yet solved, \cite{BB}.

The technique for getting MUBs for $p$  prime was  given by Schwinger, \cite{Sc}, who made use of the  properties of the Heisenberg-Weyl group. Explicit constructions in dimensions one to five   can be found in many papers, see e.g. \cite{WF}, \cite{Co}, \cite{BWB}.

We are interested in MUBs that depend on arbitrary phases because we want to use them for the construction of new complex Hadamard matrices in dimension $d=8$. To our knowledge the first who obtained such MUBs in dimension $4$ was Zauner \cite{Za}. Similar results are also given  in \cite{BWB}. 

The paper is organized as follows. In Section 2 we present some doubling formulas starting with those found by Sylvester, \cite{Sy}, and the nonlinear form of our formula, \cite{D}. 
In Section 3 we show how to construct new eight-dimensional complex Hadamard matrices that depend on five independent parameters by using our nonlinear formula.

In Section 4 we discuss the Haagerup form of the equivalence of complex Hadamard matrices, \cite{Ha}, and show that behind it stand the orthogonality relations that lead to polygons in the complex plane. Because the unitary matrices are a particular case of normal operators we propose their characteristic polynomial as the true equivalence principle.

In Section 5 we collect our results  written in the standard form introduced in \cite{TZ}.
We close the paper with a few concluding remarks.
\section{Orthogonality and doubling formulas}

The orthogonality is an important concept that appears in many mathematically, physically or engineering contexts, and it is strongly related to the angle between two objects. The simplest case is that of a vector space where the angle is found from  the usual inner product between two vectors, and if this product is zero the vectors are said to be orthogonal. This notion was extended to matrices, orthogonal designs, etc, and it is a strong tool used in many applications.  One important problem is the use of an orthogonal object for the construction of a similar one whose dimension is bigger than that of the object one started with. It seems that one of  the first who used this procedure was Sylvester in his paper, \cite{Sy}, for doubling the size of Sylvester-Hadamard matrices.

 An important observation in \cite{Sy} is the following: {\it the known theorems relating to the form of the products of two sums of 2, or of 4, or of 8 squares must exhibit instances of orthogonal matrices of this nature}. In this context he observed that the  entry wise product of the two-dimensional matrices
\begin{eqnarray}\begin{array}{cc}
\left[\begin{array}{lr}
a&b\\
b&a\end{array}\right],~\;\;&\left[\begin{array}
{cr}1&1\\
1&-1\end{array}\right]\end{array}\label{or1}\end{eqnarray} leads to the orthogonal matrix
\begin{eqnarray}
\left[\begin{array}{lr}
a&b\\
b&-a\end{array}\right]\label{or2}\end{eqnarray}
The first  matrix entering (\ref{or1}) is the simplest circulant matrix, and the second one is the core of almost all  constructions involving  Sylvester-Hadamard (real) matrices. In the four-dimensional case    the three matrices have the form
\begin{eqnarray}\begin{array}{ccc}
\left[\begin{array}{rrrr}
a&b&c&d\\
b&a&d&c\\
c&d&a&b\\
d&c&b&a\end{array}\right],~\;\;\;&\left[\begin{array}{rrrr}
1&1&1&1\\
1&-1&1&-1\\
1&-1&-1&1\\
1&1&-1&-1\end{array}\right],~\;\;&W=\left[\begin{array}{rrrr}
a&b&c&d\\
b&-a&d&-c\\
c&-d&-a&b\\
d&c&-b&-a\end{array}\right]
\end{array}\label{J1}\end{eqnarray}
The first matrix entering (\ref{J1}) is also circulant being generated by the abelian group {\bf $G=\mathbb{Z}_2\times \mathbb{Z}_2 $}, and the last matrix is the Williamson matrix, both of them being discovered {\it avant  la lettre}.

 The parameters $a,\;b,\;c,\; d$ could be square $(-1,1)$ matrices of order $m$ such that for every pair $X,Y \in (a,b,c,d)$ the product $X Y^t$ is symmetric, i.e. $XY^t=YX^t=(XY^t)^t$, where $t$ means transpose.  If they satisfy the relation
\begin{eqnarray} a a^t+b b^t+c c^t+d d^t= 4m I_{m}\end{eqnarray}
where $I_m$ is the $m$-dimensional unit matrix,  the above $W$ matrix is Hadamard of order $4m$, \cite{Wa}.
This is the usual starting point for construction of real Sylvester-Hadamard matrices by using Williamson matrix. 

 For completitude we give also his solution for $n=8$.
\begin{eqnarray}
S =\left[\begin{array}{crrrrrrr}
a&b&c&d&l&m&n&p\\
b&-a&-d&c&m&-l&p&-n\\
c&d&-a&-b&n&-p&-l&m\\
d&-c&b&-a&p&n&-m&-l\\
l&-m&-n&-p&-a&b&c&d\\
m&l&p&-n&-b&-a&d&-c\\
n&-p&l&m&-c&-d&-a&b\\
p&n&-m&l&-d&c&-b&-a\end{array}\right]\label{J2}\end{eqnarray}

A little different formula for doubling the size of a unitary matrix was found by us in paper \cite{D}  which  says that given  three $d \times d$ unitary matrices, $A,\;B,\;C$,  the 
$2 d\times 2d $ matrix
\begin{eqnarray}
\mathcal{D}=\frac{1}{\sqrt{2}}\left[\begin{array}{cc}
A&B\\
C&-CA^*B\end{array}\right]
\label{con}\end{eqnarray}
 is also unitary. Until now we used this formula only in the particular cases  $B =A$ and $C=A$ for the construction of complex Hadamard matrices, i.e. we uncovered  other two elementary block matrices similar to (\ref{or2})
\begin{eqnarray}\begin{array}{cc}
\left[\begin{array}{rr}
A&B\\
A&-B\end{array}\right]\;\;\; {\rm and}&\;\;
\left[\begin{array}{rr}
A&A\\
B&-B\end{array}\right]\end{array} \label{J3}\end{eqnarray}
 The doubling formula, i.e. the analog of Williamson matrix (\ref{J1}), has a similar form
\begin{eqnarray}
\mathcal{D}_1=\left[\begin{array}{rrrr}
A&B&C&D\\
A&-B&C&-D\\
A&B&-C&-D\\
A&-B&-C&D\end{array}\right]\label{J4}\end{eqnarray}
and so on.
 
Another important point is that we can merge the above two constructions, (\ref{or2}) and (\ref{J3}), and obtain ``hybrid'' matrices of the form
\begin{eqnarray}\begin{array}{cc}
\mathcal{D}_2=\left[\begin{array}{rrrr}
A&B&C&D\\
A&-B&C&-D\\
C&D&-A&-B\\
C&-D&-A&B\end{array}\right],&\;\;\;\;\mathcal{D}_3=\left[\begin{array}{rrrr}
A&B&A&B\\
B&-A&B&-A\\
C&D&-C&-D\\
C&-D&-C&D\end{array}\right]\end{array} \label{J5}\end{eqnarray}
which are orthogonal if  $A,\;B,\;C,\;D$ are circulant, or in a more general case
if all  $X Y^*$, $X,\;Y\in (A,\;B,\;C,\;D )$ are self-adjoint matrices, i.e.
$ X Y^*= Y X^*=( X Y^*)^* $.
The above three matrices could be also useful for finding real Sylvester-Hadamard  matrices.

In the following we do not explore these possibilities but  we make use of formula (\ref{con}) for obtaining new complex Hadamard matrices that could depend on more independent parameters than those involved in $A,\;B,$ and $C$ matrices. Here we restrict to a particular case,  that of  $A,\;B,\; C$ coming from a set  of MUBs, because it leads to a rich family of complex Hadamard matrices with a beautiful and unexpected symmetry. The first non trivial and interesting case is obtained when the three matrices have dimension four such that in this  paper we discuss this case. However we start with the  $d=2$ case to see how the method works.

\section{New eight-dimensional complex Hadamard matrices}

In the $d=2$ case  there is only one Sylvester-Hadamard matrix, up to equivalence, however we use this equivalence for taking a slightly different form for $A,\;B,\;C$ matrices. Our choice was
\begin{eqnarray}
\begin{array}{ccc}
A=\frac{1}{\sqrt{2}}\left[\begin{array}{cr}
1&1\\
1&-1\end{array}\right],\;\;B=\frac{1}{\sqrt{2}}\left[\begin{array}{cr}
1&1\\
i&-i\end{array}\right],\;\;C=\frac{1}{\sqrt{2}}\left[\begin{array}{cr}
1&i\\
1&-i\end{array}\right]\end{array} \end{eqnarray}
where in the following $i=\sqrt{-1}$. The above matrices have been multiplied by different two-dimensional diagonal phase matrices, and in this way each entry of $CA^*B$ matrix contains  four terms. The resulting matrix gets unitary by choosing a certain relationship between the arbitrary phases, and this one can be done in four different ways, and all lead, up to a factor $1/2$, to a matrix equivalent to the known result
\begin{eqnarray}
H_4(a)=\left[\begin{array}{crrr}
1&1&1&1\\
1&i\,e^{i a}&-i\,e^{i a}&-1\\
1&-i\,e^{i a}&i\,e^{i a}&-1\\
1&-1&-1&1\end{array}\right],\;\; a\in\mathbb{R}\label{had}\end{eqnarray}
By the  parametrization change,  $a\rightarrow a+\pi/2$, the matrix (\ref{had}) gets equivalent  to  the  first complex  Hadamard matrix  that depends on an arbitrary phase, result obtained in \cite{H}.

In our case, $d=4$, we use  the three mutually unbiased matrices from \cite{Za}, in different combinations, and also the above result (\ref{had}). These three matrices depend on arbitrary (real) phases and have the form

\begin{eqnarray}
\begin{array}{cc}
E_1(x)=\frac{1}{2}\left[\begin{array}{cccc}
1&1&1&1\\
1&-1&e^{i x}&-e^{i x}\\
1&1&-1&-1\\
1&-1&-e^{i x}&e^{i x}\end{array}\right],&E_2(y,z)=\frac{1}{2}\left[\begin{array}{cccc}
1&1&1&1\\
e^{i y}&-e^{i y}&e^{i z}&-e^{i z}\\
1&1&-1&-1\\
-e^{i y}&e^{i y}&e^{i z}&-e^{i z}\end{array}\right],\nonumber\end{array}
\\[3pt] 
E_3(w,z)=E_1(x)^{-1}E_2(y,z)=\frac{1}{2}\left[\begin{array}{cccc}
1&1&e^{i z}&-e^{i z}\\
1&1&-e^{i z}&e^{i z}\\
e^{i w}&-e^{i w}&1&1\\
-e^{i w}&e^{i w}&1&1\end{array}\right]~~~~~~~~~~~
\label{Z1}
\end{eqnarray}
where $w=y-x$ in the original definition, see  \cite{Za}, but in the following we consider it an independent parameter. Our first choice is to  take $A=E_1$, $B=E_2$ and $C=E_3$ and, after their multiplication   at left, and/or right by diagonal phase matrices, each entry of $CA^*B$ matrix contains  $2^4=16$ terms. Because $A,\;B,\;C$ matrices are Hadamard, the first problem to solve is the finding of those linear combinations of  arbitrary phases
such that the  $CA^*B$ matrix should also be a complex Hadamard matrix. Each entry  containing sixteen terms, the choice of independent phases is not unique, such that we can obtain many nonequivalent matrices. After that we have to check (impose) the Hadamard property for the resulting $\mathcal{D}$ matrix (\ref{con}).

 In this paper all the eight-dimensional matrices are defined up to the numerical factor $1/2 \sqrt{2}$. If in the case $d=2$ only one constraint between arbitrary phases was sufficient, in the $d=4$ case  the procedure implied at least four constraints, and the results are as follows:

\begin{eqnarray}
D_{8A}^{(5)}(a,b,c,d,f)=\left[\begin{array}{cccccccr}
1&1&1&1&1&1&1&1\\
1&e^{ia} &e^{if} &e^{id} &-e^{id} &-e^{if}&- e^{ia} &-1\\
1&e^{i b}&-e^{if} &-e^{i(b+d-a)}&e^{i(b+d-a)} &e^{if} &-e^{i b} &-1\\
1&e^{i c} &-e^{i c}&-1&-1&-e^{i c} &e^{i c} &1\\
1&-e^{i c}&e^{i c} &-1&-1&e^{i c} &-e^{i c} &1\\
1&-e^{i b}&-e^{if} &e^{i(b+d-a)}&-e^{i(b+d-a)} &e^{if} &e^{i b} &-1\\
1&-e^{ia} &e^{if} &-e^{id} &e^{id} &-e^{if}& e^{ia} &-1\\
1&-1&-1&1&1&-1&-1&1
\end{array}\right]\label{D1}
\end{eqnarray}
where the upper index,  $(5)$, means the number of independent phases, its lower  part $8A$ means its dimension, and the first such a matrix.
The above matrix  has a beautiful symmetry that can be seen by looking at the last row and column and at the $2 \times 2$ inner sub-matrix
\begin{eqnarray}\left[\begin{array}{cc}
-1&-1\\
-1&-1\end{array}\right]\nonumber\end{eqnarray}
such that both these properties can be considered as a fingerprint of such matrices. We propose that the above form should be  the standard form of a jacket matrix;  more information about jacket matrices can be found in  paper \cite{Le}.

By using the symmetry of such matrices we can bring them to the above form, (\ref {D1}), with a plus sign in front of the  first three exponential functions from  the second column and row, and we consider it a manner to present this symmetry. 
However there are some cases when the used procedure leads to matrices whose second row has a little different form, and such a matrix is the following

\begin{eqnarray}
D_{8B}^{(5)}(a,b,c,d,f)=\left[\begin{array}{cccccccr}
1&1&1&1&1&1&1&1\\
1&e^{ia} &-e^{ia} &e^{id} &-e^{id} &-e^{ia}& e^{ia} &-1\\
1&e^{i b}&e^{i(b-c+f)} &-e^{id}&e^{id} &-e^{i(b-c+f)} &-e^{i b} &-1\\
1&e^{i c} &-e^{i f}&-1&-1&e^{i f} &-e^{i c} &1\\
1&-e^{i c}&e^{i f} &-1&-1&-e^{i f} &e^{i c} &1\\
1&-e^{i b}&-e^{i(b-c+f)} &-e^{id}&e^{id} &e^{i(b-c+f)} &e^{i b} &-1\\
1&-e^{ia} &e^{ia} &e^{id} &-e^{id} &e^{ia}&- e^{ia} &-1\\
1&-1&-1&1&1&-1&-1&1
\end{array}\right]\label{D2}
\end{eqnarray}
Looking carefully at $D_{8B}^{(5)}$ matrix one can observe that it can be brought to the same form as $D_{8A}^{(5)}$ matrix if we permute between themselves the fourth and fifth columns, followed by a permutation of  the second and the third rows, and respectively of the sixth and seventh ones. If on this new form we make simultaneously  the changes $a\rightarrow b$, and  $b\rightarrow a$, we find the matrix
\begin{eqnarray}
D_{8C}^{(5)}(a,b,c,d,f)=\left[\begin{array}{cccccccr}
1&1&1&1&1&1&1&1\\
1&e^{ia} &e^{i(a-c+f)} &e^{id}&-e^{id} &-e^{i(a-c+f)} &-e^{i a} &-1\\
1&e^{ib} &-e^{ib} &-e^{id} &e^{id} &-e^{ib}& e^{ib} &-1\\
1&e^{i c} &-e^{i f}&-1&-1&e^{i f} &-e^{i c} &1\\
1&-e^{i c}&e^{i f} &-1&-1&-e^{i f} &e^{i c} &1\\
1&-e^{ib} &e^{ib} &-e^{id} &e^{id} &e^{ib}&- e^{ib} &-1\\
1&-e^{ia} &-e^{i(a-c+f)} &e^{id}&-e^{id} &e^{i(a-c+f)} &e^{i a} &-1\\
1&-1&-1&1&1&-1&-1&1
\end{array}\right]\label{D3}
\end{eqnarray}
Although the above matrix was obtained by using  the usual definition  of   matrix equivalence for complex Hadamard matrices, \cite{Ha}, $D_{8B}^{(5)}$ and  $D_{8C}^{(5)}$ are not equivalent. The columns(rows) permutation, \textit{together with}   ``benign'' re-parametrization of the form, $a\rightarrow b$, and  $b\rightarrow a$, lead to new non equivalent matrices.

By using equivalent matrices to $E_1,\;E_2,\;E_3$ for $A,\;B,\;C$, obtained by permutations of rows/columns one get  similar matrices, that  are  not equivalent to (\ref{D1}), (\ref{D2}), or (\ref{D3})
\begin{eqnarray}
D_{8D}^{(5)}(a,b,c,d,f)=\left[\begin{array}{cccccccr}
1&1&1&1&1&1&1&1\\
1&e^{ia} &e^{if} &e^{id} &-e^{id} &-e^{if}& -e^{ia} &-1\\
1&e^{i b}&-e^{ib} &-e^{id}&e^{id} &-e^{ib} &e^{i b} &-1\\
1&e^{i c} &-e^{i(c+ f-a)}&-1&-1&e^{i(c+ f-a)} &-e^{i c} &1\\
1&-e^{i c}&e^{i(c+ f-a)} &-1&-1&-e^{i(c+ f-a)} &e^{i c} &1\\
1&-e^{i b}&e^{ib} &-e^{id}&e^{id} &e^{ib} &-e^{i b} &-1\\
1&-e^{ia} &-e^{if} &e^{id} &-e^{id} &e^{if}& e^{ia} &-1\\
1&-1&-1&1&1&-1&-1&1
\end{array}\right]\label{D4}
\end{eqnarray}

\begin{eqnarray}
D_{8E}^{(5)}(a,b,c,d,f)=\left[\begin{array}{cccccccr}
1&1&1&1&1&1&1&1\\
1&e^{ia} &e^{if} &e^{i(a-b+d)} &-e^{i(a-b+d)} &-e^{if}&- e^{ia} &-1\\
1&e^{i b}&-e^{if} &-e^{id}&e^{id} &e^{i f} &-e^{i b} &-1\\
1&e^{i c} &-e^{i c}&-1&-1&-e^{i c} & e^{i c} &1\\
1&-e^{i c}&e^{i c} &-1&-1&e^{i c} &-e^{i c} &1\\
1&-e^{i b}&-e^{if} &e^{id}&-e^{id} &e^{if} &e^{i b} &-1\\
1&-e^{ia} &e^{if} &-e^{i(a-b+d)} &e^{i(a-b+d)} &-e^{if}& e^{ia} &-1\\
1&-1&-1&1&1&-1&-1&1
\end{array}\right]\label{D5}
\end{eqnarray}
Another matrix with a little different form is 
\begin{eqnarray}
D_{8F}^{(5)}(a,b,c,d,f)=\left[\begin{array}{cccccccr}
1&1&1&1&1&1&1&1\\
1&e^{ia} &e^{if} &e^{id} &-e^{id} &-e^{if}&- e^{ia} &-1\\
1&e^{i a}&-e^{if} &-e^{id}&e^{id} &e^{i f} &-e^{i a} &-1\\
1&e^{i c} &-e^{i c}&-1&-1&-e^{i c} & e^{i c} &1\\
1&-e^{i c}&e^{i c} &-1&-1&e^{i c} &-e^{i c} &1\\
1&-e^{i a}&-e^{i(b-d+f)} &e^{ib}&-e^{ib} &e^{i(b-d+f)} &e^{i a} &-1\\
1&-e^{ia} &e^{i(b-d+f)} &-e^{ib} & e^{ib}&-e^{i(b-d+f)}& e^{ia} &-1\\
1&-1&-1&1&1&-1&-1&1
\end{array}\right]\label{D6}
\end{eqnarray}
Now we  introduce into our game the matrix $H_4(a)$, Eq.(\ref{had}), that has a special dependence  on the imaginary unit $i$ by taking one or more  of the  $A,\;B,\;C$ matrices,  equivalent  to $H_4$, each one depending on a different phase. Using the same procedure as above we got

\begin{eqnarray}
D_{8G}^{(5)}(a,b,c,d,f)=\left[\begin{array}{cccccccr}
1&1&1&1&1&1&1&1\\
1&e^{ia} &e^{if} &i e^{id} &-i  e^{id} &-e^{if}&- e^{ia} &-1\\
1&e^{i b}&-e^{if} &-i e^{i(b+d-a)}&i e^{i(b+d-a)} &e^{if} &-e^{i b} &-1\\
1&i e^{i c} &-i e^{i c}&-1&-1&-i e^{i c} &i e^{i c} &1\\
1&-i e^{i c}&i e^{i c} &-1&-1&i e^{i c} &-i e^{i c} &1\\
1&-e^{i b}&-e^{if} &i e^{i(b+d-a)} &-i e^{i(b+d-a)} &e^{if} &e^{i b} &-1\\
1&-e^{ia} &e^{if} &-i e^{id} &i e^{id} &-e^{if}& e^{ia} &-1\\
1&-1&-1&1&1&-1&-1&1
\end{array}\right]\label{D7}
\end{eqnarray}
\begin{eqnarray}
D_{8H}^{(5)}(a,b,c,d,f)=\left[\begin{array}{cccccccr}
1&1&1&1&1&1&1&1\\
1&e^{ia} &i\,e^{if} &e^{id} &-e^{id} &-i\,e^{if}&- e^{ia} &-1\\
1&e^{i b}&-i\,e^{i(b+f-a)} &-e^{id}&e^{id} &i\,e^{i(b+f-a)} &-e^{i b} &-1\\
1&e^{i c} &-e^{i c}&-1&-1&-e^{i c} &e^{i c} &1\\
1&-e^{i c}&e^{i c} &-1&-1&e^{i c} &-e^{i c} &1\\
1&-e^{i b}&i\,e^{i(b+f-a)} &-e^{id}&e^{id} &-i\,e^{i(b+f-a)} &e^{i b} &-1\\
1&-e^{ia} &-i\,e^{if} &e^{id} &-e^{id} &i\,e^{if}& e^{ia} &-1\\
1&-1&-1&1&1&-1&-1&1
\end{array}\right]\label{D8}
\end{eqnarray}
It is easily seen that by the re-parametrization $c\rightarrow c-\pi/2$ and 
 $d\rightarrow d-\pi/2$, and, respectively,  $f\rightarrow f-\pi/2$, the matrices (\ref{D7}) and (\ref{D8}) transforms into the matrix (\ref{D1}), and respectively (\ref{bas2}). The matrices with, and without $i$, are not equivalent under the usual form of equivalence, however we will put them in the same class, as we will show latter.

 Matrices with the same fingerprint as those found by us  appeared in other different contexts.
For example a similar  matrix can be recovered  from the complex Hadamard matrix, $P_8$, found  in \cite{BAS}, that originally depends on three non-null complex parameters, and by permutation of rows and columns it can be brought to the form
\begin{eqnarray}
P_{8}^{(3)}(a,b,c)=\left[\begin{array}{rrrrrrrr}
1&1&1&1&1&1&1&1\\
1&e^{i a}&e^{ib}&e^{i c}&-e^{i c} &-e^{ib}&-e^{ia}&-1\\
1&e^{i a}&-e^{ib}&-e^{i c}&e^{i c} &e^{ib}&-e^{ia}&-1\\
1&1&-1&-1&-1&-1&1&1\\
1&-1&1&-1&-1&1&-1&1\\
1&-e^{i a}&e^{ib}&-e^{i c}&e^{i c} &-e^{ib}&e^{ia}&-1\\
1&-e^{i a}&-e^{ib}&e^{i c}&-e^{i c} &e^{ib}&e^{ia}&-1\\
1&-1&-1&1&1&-1&-1&1\\
\end{array}\right]\label{bas1}\end{eqnarray}
that has the same symmetry. Thus we can  recover from it a matrix which depends on five arbitrary phases, namely

\begin{eqnarray}
D_{8I}^{(5)}(a,b,c,d,f)=\left[\begin{array}{rrcrrcrr}
1&1&1&1&1&1&1&1\\
1&e^{i a}&e^{if}&e^{i d}&-e^{i d} &-e^{if}&-e^{ia}&-1\\
1&e^{ib}&-e^{i(b+f-a)}&-e^{id}&e^{id}&e^{i(b+f-a)}&-e^{ib}&-1\\
1&e^{ic}&-e^{ic}&-1&-1&-e^{ic}&e^{ic}&1\\
1&-e^{ic}&e^{ic}&-1&-1&e^{ic}&-e^{ic}&1\\
1&-e^{ib}&e^{i(b+f-a)}&-e^{id}&e^{id}&-e^{i(b+f-a)}&e^{ib}&-1\\
1&-e^{i a}&-e^{if}&e^{i d}&-e^{i d} &e^{if}&e^{ia}&-1\\
1&-1&-1&1&1&-1&-1&1\\
\end{array}\right]\label{bas2}\end{eqnarray}
and its transpose 
 for $f=a$ and $d=0$ coincides with the matrix (\ref{bas1}). 

 Another example comes from a paper by Horadam, \cite{Ho}, who found a matrix, $K_4(i)$, that is {\em equivalent to the back circulant matrix derived from a quadriphase perfect sequence of length 8, and such sequences are rare.} 

  By permuting rows and columns it can be written as

\begin{eqnarray}
K_4(i)=\left[\begin{array}{rrrrrrrr}
1&1&1&1&1&1&1&1\\
1&1&i &i &-i  &-i &-1&-1\\
1&i &1&-i &i&-1&-i &-1\\
1&i&-i&-1&-1&-i&i&1\\
1&-i&i&-1&-1&i&-i&1\\
1&-i&-1&-i&i&1&i&-1\\
1&-1&-i&i&-i&i&1&-1\\
1&-1&-1&1&1&-1&-1&1\\
\end{array}\right]\label{kat1}\end{eqnarray}

The above matrix   has  a similar symmetry  to the previous matrices, and, by consequence, we can try to find the ``natural'' matrix it comes from. After few trials we found that the matrix (\ref{kat1}) is a particular case of the following complex Hadamard matrix that depends on five phases

\begin{eqnarray}
D_{8J}^{(5)}(a,b,c,d,f)=\left[\begin{array}{cccccccc}
1&1&1&1&1&1&1&1\\
1&e^{i a}&i e^{id}&i e^{if} & -i e^{if}&-i e^{id} &-e^{i a} &-1\\
1&i e^{ib} &e^{i(b+d-a)}&-i e^{if} &i e^{if} &-e^{i(b+d-a)} &-i e^{ib} &-1\\
1&i e^{ic}&-i e^{ic} &-1&-1&-i e^{ic} &i e^{ic} &1\\
 1&-i e^{ic}&i e^{ic} &-1&-1&i e^{ic} &-i e^{ic} &1 \\
  1&-i e^{ib} &-e^{i(b+d-a)}&-i e^{if} &i e^{if} &e^{i(b+d-a)} &i e^{ib} &-1\\ 
 1&-e^{i a}&-i e^{id}&i e^{if} &- i e^{if}&i e^{id} &e^{i a} &-1 \\
1&-1&-1&1&1&-1&-1&1\\
\end{array}\right]\label{kat2}\end{eqnarray}
If one makes the phases change $b\rightarrow b-\pi/2,\;c\rightarrow c-\pi/2,\; d\rightarrow d-\pi/2,\;f\rightarrow f-\pi/2$ one gets
\begin{eqnarray}
D_{8K}^{(5)}(a,b,c,d,f)=\left[\begin{array}{cccccccc}
1&1&1&1&1&1&1&1\\
1&e^{i a}& e^{id}& e^{if} & - e^{if}&- e^{id} &-e^{i a} &-1\\
1& e^{ib} &-e^{i(b+d-a)}& -e^{if} & e^{if} &e^{i(b+d-a)} &- e^{ib} &-1\\
1& e^{ic}&- e^{ic} &-1&-1&- e^{ic} & e^{ic} &1\\
 1&- e^{ic}& e^{ic} &-1&-1& e^{ic} &- e^{ic} &1 \\
  1&- e^{ib} &e^{i(b+d-a)}& -e^{if} & e^{if} &-e^{i(b+d-a)} & e^{ib} &-1\\ 
 1&-e^{i a}&- e^{id}& e^{if} & - e^{if}& e^{id} &e^{i a} &-1 \\
1&-1&-1&1&1&-1&-1&1\\
\end{array}\right]\label{kat3}\end{eqnarray}

The above matrices are new, see e.g. \cite{TZ} and \cite{BTZ}, where the only known result parametrized by five phases is the complex Hadamard matrix stemming from the Fourier matrix $F_8$.

All these matrices  generate real and complex Hadamard matrices   that depend on $\pm 1$ and $\pm i$ when the parameters $(a,b,c,d,f)$ takes 0, $\pm\pi/2$ or $\pi$ values. There are 32 such matrices and the number of  $i$ entering each one is quantized $q=8\,n,\;\,n=0,\,1,\,2,\,3,\,4$.  If we consider the matrix $D_{8A}^{(5)}(a,b,c,d,f)$, then there is one real matrix for $n=0$, obtained when all the five parameters take $\pi/2$ value, eight matrices when $n=1$, fourteen for   $n=2$, eight for $n=3$, and again  one when  $n=4$.

In the same time all the matrices generate 31 complex Hadamard matrices   that depend on five arbitrary phases by doing re-parametrization of the form $a\rightarrow a + \pi/2$, where $a$ denotes any phase entering a given matrix.

One easily obtain similar matrices depending on {\it four} independent phases, the simplest method being to make zero a linear combination with integer coefficients of the five independent phases. However we used the same method by choosing other combinations of arbitrary phases entering the matrix $C^*AB$ and a few results are the following  

\begin{eqnarray}
D_{8A}^{(4)}(a,b,c,d)=\left[\begin{array}{cccccccr}
1&1&1&1&1&1&1&1\\
1&e^{ia} &-e^{ia} &e^{id} &-e^{id} &-e^{ia}& e^{ia} &-1\\
1&e^{i b}&e^{ia} &-e^{id}&e^{id} &-e^{ia} &-e^{i b} &-1\\
1&e^{i c} &-e^{i(a-b+ c)}&-1&-1&e^{i(a-b+ c)} &-e^{i c} &1\\
1&-e^{i c}&e^{i(a-b+ c)} &-1&-1&-e^{i (a-b+ c) } &e^{i c} &1\\
1&-e^{i b}&-e^{ia} &-e^{id}&e^{id} &e^{ia} &e^{i b} &-1\\
1&-e^{ia} &e^{ia} &e^{id} &-e^{id} &e^{ia}& -e^{ia} &-1\\
1&-1&-1&1&1&-1&-1&1
\end{array}\right]\label{D10}
\end{eqnarray}

\begin{eqnarray}
D_{8B}^{(4)}(a,b,c,d)=\left[\begin{array}{cccccccr}
1&1&1&1&1&1&1&1\\
1&e^{ia} &-e^{ia} &e^{id} &-e^{id} &-e^{ia}& e^{ia} &-1\\
1&e^{i b}&e^{i(b-a)} &-e^{id}&e^{id} &-e^{i(b-a)} &-e^{i b} &-1\\
1&e^{i c} &-e^{i(c-a)}&-1&-1&e^{i( c-a)} &-e^{i c} &1\\
1&-e^{i c}&e^{i( c-a)} &-1&-1&-e^{i ( c-a) } &e^{i c} &1\\
1&-e^{i b}&-e^{i(b-a)} &-e^{id}&e^{id} &e^{i(b-a)} &e^{i b} &-1\\
1&-e^{ia} &e^{ia} &e^{id} &-e^{id} &e^{ia}& -e^{ia} &-1\\
1&-1&-1&1&1&-1&-1&1
\end{array}\right]\label{D11}
\end{eqnarray}

The most interesting cases seem to be
\begin{eqnarray}
D_{8C}^{(4)}(a,b,c,d)=\left[\begin{array}{cccccccr}
1&1&1&1&1&1&1&1\\
1&e^{ia} & e^{id} &i e^{i(a-c+d)}&-ie^{i(a-c+d)}  &-e^{id}& -e^{ia} &-1\\
1&e^{i b}&-e^{id} &-i e^{i(b-c+d)}&i e^{i(b-c+d)} &e^{id} &-e^{i b} &-1\\
1&e^{i c} &-e^{ic}&-1&-1&-e^{ic} &e^{i c} &1\\
1&-e^{i c}&e^{ic} &-1&-1&e^{i  c } &-e^{i c} &1\\
1&-e^{i b}&-e^{id} &i e^{i(b-c+d)}&-i e^{i(b-c+d)} &e^{id} &e^{i b} &-1\\
1&-e^{ia} &e^{id} &-i e^{i(a-c+d)} &ie^{i(a-c+d)}  & -e^{id}& e^{ia} &-1\\
1&-1&-1&1&1&-1&-1&1
\end{array}\right]\label{D12}
\end{eqnarray}
\begin{eqnarray}
D_{8D}^{(4)}(a,b,c,d)=\left[\begin{array}{cccccccr}
1&1&1&1&1&1&1&1\\
1&e^{ia} & i e^{i(c+d-a)} & e^{id}&-e^{id}  &-i e^{i(c+d-a)}& -e^{ia} &-1\\
1&e^{i b}&-i e^{i(c+d-a)} &- e^{i(b+d-a)}&e^{i(b+d-a)} & i e^{i(c+d-a)} &-e^{i b} &-1\\
1&e^{i c} &-e^{ic}&-1&-1&-e^{ic} &e^{i c} &1\\
1&-e^{i c}&e^{ic} &-1&-1&e^{i  c } &-e^{i c} &1\\
1&-e^{i b}&-i e^{i(c+d-a)} & e^{i(b+d-a)}&- e^{i(b+d-a)} & i e^{i(c+d-a)} &e^{i b} &-1\\
1&-e^{ia} &i e^{i(c+d-a)} &- e^{id} &e^{id}  & -i e^{i(c+d-a)} & e^{ia} &-1\\
1&-1&-1&1&1&-1&-1&1
\end{array}\right]\label{D13}
\end{eqnarray}
 because they cannot be brought by re-parametrization  to matrices similar to  (\ref{D10}) and  (\ref{D11}). This shows the importance of the $H_4(a)$ matrix  which allows to obtain more nonequivalent matrices that depend on four parameters.  Thus the nonlinear term $C A^* B$, whose each entry depends on 16 terms (phases), leads to non intersecting orbits depending on five, four or three phases, such that there do exist Hadamard matrices that depend on four or three phases that cannot be obtained from the generic five-dependent ones, e.g. $D_{8A}^{(5)}$, by collapsing some parameters, or by re-parametrization of the form  $a\rightarrow a\pm\pi/2$, where $a$ denotes any parameter entering a given matrix.
The above four matrices depending on four phases are new. In the known catalogue, \cite{BTZ}, there is only one such  matrix, that found in paper \cite{MRS}, which  has another symmetry, showing that the set of all eight-dimensional complex Hadamard matrices has a rich structure, and the results obtained till now, including these ones, represent  only  partial  results.

An other example with three parameters is the following
\begin{eqnarray}
D_{8A}^{(3)}(a,b,c)=\left[\begin{array}{cccccccr}
1&1&1&1&1&1&1&1\\
1& e^{ia} & e^{ia} & ie^{ic} & -i e^{ic} &- e^{ia}& - e^{ia} &-1\\
1&e^{i a}&-e^{ia} &-ie^{ic}&ie^{ic} &-e^{ia} &e^{i a} &-1\\
1&e^{i b} &- e^{ib}&-1&-1& e^{ib} &-e^{i b} &1\\
1&-e^{i b}& e^{ib} &-1&-1&- e^{i b } &e^{i b} &1\\
1&-e^{i a}&-i e^{ia} &-e^{ic}&e^{ic} &e^{ia} &i e^{i a} &-1\\
1&- e^{ia} &i e^{ia} &e^{ic} &-e^{ic}&  e^{ia}&-i  e^{ia} &-1\\
1&-1&-1&1&1&-1&-1&1
\end{array}\right]\label{D14}
\end{eqnarray}

\section{Equivalence of complex Hadamard matrices}

The present day used form for the equivalence of complex Hadamard matrices is essentially the same as that for real matrices,  and can be stated as follows. Two  Hadamard matrices $H_1$ and $H_2$ are called equivalent, written  $H_1\approx H_2$, if there exist two diagonal unitary matrices $D_1$ and $D_2$ and permutation matrices $P_1$ and $P_2$ such that 
\begin{eqnarray}
H_1\,=\, D_1P_1H_2P_2 D_2\label{ha1}\end{eqnarray}
The above relation allows  to bring any Hadamard matrix two its Sylvester form,
\cite{Sy}, i.e. to a matrix whose all entries from the first row and column are equal to unity.

For complex matrices Haagerup suggested, \cite{Ha}, as a  practical method for detecting (non)equivalence of two Hadamard matrices in more complicated cases,  the use of the set of numbers
\begin{eqnarray} \{H_{ij} H_{kl}\bar{ H}_{il}\bar{H}_{kj}\;\;|\;\;i,j,k,l\,=\,1,\dots,n\}\label{ph}\end{eqnarray}
 for both the matrices $H_1$ and  $H_2$, where bar denote complex conjugation and $H_{ij}$
 the entries of the matrix $H$, because the above product is invariant under equivalence (\ref{ha1}). If the two sets of phases (\ref{ph}) coincide  the matrices are equivalent.
 In some sense (\ref{ph})  is a local property that does not take into account the global properties of a complex Hadamard matrix. Let see what is behind it.

For the first time such a product appeared in the paper \cite{Ja}, see also \cite{JS}, which deals with the structure of the  $3\times 3$ quark mass matrices in the standard electroweak model. These matrices for up and down quarks are diagonalized by unitary matrices $U$ and $U'$ and their product $V=U\,U'^*$ defines the so called Kobayashi-Maskawa unitary matrix. Jarlskog showed that the generic products {\it Im}$(V_{ij} V_{kl} \bar{V}_{il} \bar{V}_{kj} )$ are invariant phases of the quark mixing  matrix. In fact the invariant phases are directly related to the angles of the so called unitarity triangles, see e.g. \cite{D1}. 

In the general case the unitarity polygons are generated by the rows and columns orthogonality. For example the orthogonality relation of  $j$ and $l$ columns is written as 
\begin{eqnarray}\sum_{i=1}^n H_{ij}\bar{ H}_{il}=0\label{orth1}\end{eqnarray}
and if we consider its entries as vectors in a two-dimensional space the above relation, (\ref{orth1}), defines a polygon in the complex plane. Because for an arbitrary $n \times n$ unitary matrix all the terms entering (\ref{orth1}) could have different moduli  the physicists divided the above relation by one term, let say $H_{kj}\bar{H}_{kl}$, such that one leg length is unity, and (\ref{orth1}) has the new form
\begin{eqnarray}1+\sum_{i=1,i\ne k}^n\frac{ H_{ij}\bar{ H}_{il}}{H_{kj}\bar{H}_{kl}}=0\label{orth2}\end{eqnarray}
For Hadamard matrices all legs length is unity if we redefine them without the factor $1/\sqrt{n}$,  and then the  generic term entering the above sum is $H_{ij}H_{kl}\bar{H}_{kj}\bar{ H}_{il}$ which coincides with (\ref{ph}). Thus  for complex Hadamard matrices the orthogonal relations (\ref{orth1}) generate many  polygons in the complex plane  whose maximal dimension is $n$. All these polygons coming from all orthogonality relations define the fingerprint of a given complex Hadamard matrix, and we could say that two complex Hadamard matrices are equivalent if their fingerprints coincide. This fingerprint gives a global description of matrix properties. 
To see how this equivalence recipe works let us consider the matrix $H_4(a)$ for $a=0.$
\begin{eqnarray}
H_4(0)=\left[\begin{array}{crrr}
1&1&1&1\\
1&i&-i&-1\\
1&-i&i&-1\\
1&-1&-1&1\end{array}\right]\label{had1}\end{eqnarray}
It is easily seen that the fingerprint of the above matrix consists in the point $O(0,0)$ of the coordinate axes,  and the square $OABC$ defined by $A=(1,0),\,B=(1,1),\,C=(0,1)$. When  $a\ne 0$ the  $H_4(a)$ fingerprint
is the same point $O$, and a rhomb whose precise shape depends on the phase $a$. In conclusion we have a continuum of independent Hadamard matrices, all of them  having the classical Hadamard form \cite{H} because the factor $i$ can be absorbed in the phase $a$.

We consider now two six-dimensional complex Hadamard matrices
\begin{eqnarray}\begin{array}{cc}
D_{6A}=\left[\begin{array}{rrrrrrrr}
1&1&1&1&1&1\\
1&-1&-1 &1 &i &-i\\
1&-1 &-i &-1&1&i\\
1&1&-1&-i&-1&i\\
1&i&1&-1&-1&-i\\
1&-i&i&i&-i&-1\\
\end{array}\right],&
D_{6B} =\left[\begin{array}{rrrrrrrr}
1&1&1&1&1&1\\
1&-1&i &i &-i  &-i \\
1&-i &-1&1 &-1&i\\
1&-i&1&-1&i&-1\\
1&i&-1&-i&1&-1\\
1&i&-i&-1&-1&1\\
\end{array}\right]\end{array}\label{hH1}\end{eqnarray}
According to relation (\ref{ph})  the above two matrices must be equivalent since both of them have the same number of $i$, namely 12,  and the same number of  $-1$, namely  9. However they cannot be equivalent because $D_1$ is a symmetric matrix, while  $D_2$ is self-adjoint.
The relations (\ref{orth1}) lead  to similar fingerprints for both the matrices which consists in the preceding square $OABC$ and the rectangle  $OAB_1C_1$ with $B_1=(1,2),\,C_1=(0,2)$,  and the rectangle $OA_1B_2C$ defined by  $A_1=(0,2),\,B_2=(2,1)$, and it is difficult to see if    $D_{6A}$ and  $D_{6B}$ are (non)equivalent matrices. In fact we have to compare all the $6\times 5=30$ polygons generated by  $D_{6A}$ orthogonality,  and, respectively, by  $D_{6B}$.

In the case of matrices $D_{8*}^{(5)}(a,b,c,d,f)$, $*=A,\dots, L$, the polygons can start with the point $O$, ending with octagons. For example the matrices $D_{8B}^{(5)}(a,b,c,d,f)$ and $D_{8C}^{(5)}(a,b,c,d,f)$ are not equivalent because the octagons defined by the orthogonality of the first and third columns have different shapes, etc. However, like the case of $H_4(a)$ matrix, all matrices whose some entries contain products of the imaginary unit with arbitrary phases, and in case the imaginary unit can be absorbed by re-parametrization, will be considered as defining one continuous class.

Defining  the fingerprint class  by all the orthogonality relations of the form (\ref{orth1}) has an advantage over the present day used form  because it implies that the matrices $H$, $H^t$ and $\bar{H}$ are all equivalent by reducing the number of non equivalent matrices.

In the mathematical literature there is an other type of matrix equivalence for the class of normal matrices.  A matrix $N$ is normal if it commutes with its adjoint $N^*$, i.e., $N N^*=N^*N$, for which the equivalence has a simple form. Every normal matrix is similar to a diagonal matrix $D$, which means that there exists a unitary matrix $U$ such that $N=U\,D\,U^*$, \cite{Pu}, p. 357, or \cite{Kel}.
 Two important classes of normal matrices  are the unitary and self-adjoint matrices such that we can say that two matrices, $A_1$ and $A_2$,  are equivalent if and only if they have the same spectrum, or equivalently, the characteristic polynomials satisfy the relation, $det(\lambda I_n- A_1)= det(\lambda I_n- A_2)$, up to a multiplicative numerical factor, where $det$ means the determinant of the corresponding matrix.

For example  the spectra of  $D_{6A}$ and $D_{6B}$ matrices are as follows
\begin{eqnarray}
Sp(D_{6A})&=&\left( -1,1,\frac{1}{6}i(-\sqrt{6} \pm i\sqrt{30}),\frac{1}{6}(-\sqrt{6}\pm i\sqrt{30}  \right)\\
Sp(D_{6B})&=&\left(-1,-1,-1,1,1,1\right)\end{eqnarray}
such that the above matrices are not equivalent.

The circulant matrix generated by the row 
$(1,i d,-d,-i,-1/d,i/d),$ where $d$ is one solution of the equation $d^2-(1-\sqrt{3})d+1=0$, see \cite{BF}, or the matrix (2.26) in \cite{Ha},
has the same spectrum as $D_{6B}$, so they are equivalent.

In general the spectrum of a matrix cannot be explicitly found, however its characteristic polynomial can be, so the (non)equivalence can be established in most the cases.

The characteristic polynomials of our matrices, up to numerical factors, are as follows
\begin{eqnarray}
P_{D_{8A}^{(5)}}(\lambda)=(1-\lambda^2)^2\left(2\,\lambda^4- \sqrt{2}\,   e^{ ia}\lambda^3+ \sqrt{2} \,  e^{ i(-a+b+c+d+f)}\lambda-2\, e^{ i(b+c+d+f)}\right)\label{8A}
\end{eqnarray}
 If one makes the re-parametrization $d\rightarrow d+\pi$, which is equivalent to the  interchange of the fourth and five columns of the  $D_{8A}^{(5)} $ matrix, one finds
\begin{eqnarray}
P_{D_{8A}^{(5!)}}(\lambda)&=&(1-\lambda^2)^2\left(2\,\lambda^4- \sqrt{2}\,   e^{ ia}\lambda^3- \sqrt{2} \,  e^{ i(-a+b+c+d+f)}\lambda+2\, e^{ i(b+c+d+f)}\right)\label{8A!}
\end{eqnarray}
thus one can say that the corresponding matrices $D_{8A}^{(5)}  $ and  $D_{8A}^{(5!)} $ are {\it not} equivalent. The same form for the relation (\ref{8A!}) can be obtained by using a re-parametrization with any other parameter, $b,\,c,$ or $f$, e.g., $b \to b+\pi, $  and all the four such obtained matrices are equivalent, as it easily seen from relations (\ref{8A}) and (\ref{8A!}).

 For the next four matrices the polynomials are

\begin{eqnarray}
P_{D_{8B}^{(5)}}(\lambda) &=&(1-\lambda^2)^2\left(2\, \lambda^4- \sqrt{2}\,   e^{i(b-c+f)}\lambda^3- \sqrt{2} \,  e^{ i(a+ c+d)}\lambda+2\, e^{ i(a+b+d+f)}\right)\label{8B}\\
P_{D_{8C}^{(5)}}(\lambda)  &=&(1-\lambda^2)^2\left(2\,\lambda^4- \sqrt{2}\,   e^{ia}\lambda^3+ \sqrt{2} \,  e^{ i(b+d+f)}\lambda-2\, e^{ i(a+b+d+f)}\right)\label{8C}\\
P_{D_{8D}^{(5)}}(\lambda) &=&(1-\lambda^2)^2\left(2 \lambda^4- \sqrt{2}\,   e^{ia}\lambda^3+ \sqrt{2} \,  e^{ i(-a+b+c+d+f)}\lambda-2\, e^{ i(b+c+d+f)}\right)\label{8D}\\
P_{D_{8E}^{(5)}}(\lambda)  &=&(1-\lambda^2)^2\left(2\,\lambda^4- \sqrt{2}\,e^{ia}\lambda^3+ \sqrt{2} \,e^{ i( c+d+f)}\lambda-2\, e^{ i(a+c+d+f)}\right)\label{8E}
\end{eqnarray}
The following matrix, $D_{8F}^{(5)}$, leads to a more complicated polynomial
\begin{eqnarray}
P_{D_{8F}^{(5)}}(\lambda) &=&(1-\lambda^2)^2\left(4\,\lambda^4- \sqrt{2}\,(2\,e^{ia}+e^{i(b-d+f)}-e^{i f})\lambda^3+ 2(e^{i(b-d)}-1)(e^{ i(c+d)}+e^{ i( a+f)})\lambda^2\right.\nonumber\\&&\left.-\sqrt{2}\,e^{ic}( e^{ i(a+b)}-e^{ i(a+d)}-2\,e^{ i(b+f)})\lambda-4e^{ i(a+b+c+f)}\right)\label{8F}\end{eqnarray}
The next matrix, $D_{8G}^{(5)}$, is interesting because  by the re-parametrization $c\rightarrow c -\pi/2,\;{\rm and}\; d\rightarrow c -\pi/2 $ it is transformed into the matrix  $D_{8A}^{(5)}$.  Its polynomial has the form
\begin{eqnarray}
P_{D_{8G}^{(5)}}(\lambda) =(1-\lambda^2)^2\left(2\,\lambda^4- \sqrt{2}\,   e^{ ia}\lambda^3- \sqrt{2} \,  e^{ i(-a+b+c+d+f)}\lambda+2\, e^{ i(b+c+d+f)}\right)\label{8G}
\end{eqnarray}
 
The formulas (\ref{8A}) and (\ref{8G}) differ only by the sign of the last two terms, and could  show us  which are the matrices that are equivalent. In fact there are six such transformations involving any pair from the set $(b,c,d,f)$ and all these four parameters, that preserve the form (\ref{8G}). If we make the transformation $a\rightarrow a+\pi$ in matrix (\ref{D7}) one gets
\begin{eqnarray}
P_{D_{8G}^{(5!)}}(\lambda) =(1-\lambda^2)^2\left(2\,\lambda^4+ \sqrt{2}\,   e^{ ia}\lambda^3+ \sqrt{2} \,  e^{ i(-a+b+c+d+f)}\lambda+2\, e^{ i(b+c+d+f)}\right)\label{8G!}
\end{eqnarray}

The next two matrices lead to 
\begin{eqnarray}
P_{D_{8H}^{(5)}}(\lambda) &=&(1-\lambda^2)^2(e^{i(c+d)}-\lambda^2)\left(2\, \lambda^2+ \sqrt{2}\,(i\,e^{i(-a+b+f)}- e^{ ia})\lambda -2\,i\, e^{ i(b+f)}\right)\label{8H}\\
P_{D_{8I}^{(5)}}(\lambda) &=&(1-\lambda^2)^2(e^{i(c+d)}-\lambda^2)\left(2\, \lambda^2 + \sqrt{2}\,(e^{i(-a+b+f)}- e^{ia})\lambda -2\, e^{ i(b+f)}\right)\label{8I}
\end{eqnarray}
whose spectra can be computed explicitly. A similar case is 
\begin{eqnarray}
P_{D_{8J}^{(5)}}(\lambda)  &=&(1-\lambda^2)^2(e^{i(c+f)}+\lambda^2)\left(2\,\lambda^2- \sqrt{2}(e^{i(-a+b+d)} +  e^{ia})\lambda +2\, e^{ i(b+d)}\right)\label{8J}\\
 P_{D_{8K}^{(5)}}(\lambda)  &=&(1-\lambda^2)^2(e^{i(c+f)}-\lambda^2)\left(2\,\lambda^2+ \sqrt{2}\,(e^{i(-a+b+d)}- e^{ ia})\lambda -2\, e^{ i(b+d)}\right)\label{8K}
\end{eqnarray}
Matrices of the form $D_{8F}^{(5)}$ seem to be rare, another example being the
following
\begin{eqnarray}
D_{8L}^{(5)}(a,b,c,d,f)=\left[\begin{array}{cccccccr}
1&1&1&1&1&1&1&1\\
1&e^{ia} &e^{i(b+d-f)} &e^{id} &-e^{id} &- e^{i(b+d-f)}&- e^{ia} &-1\\
1&e^{i a}&-e^{i(b+d-f)} &-e^{id}&e^{id} &e^{i(b+d-f)} &-e^{i a} &-1\\
1&e^{i c} &-e^{i c}&-1&-1&-e^{i c} & e^{i c} &1\\
1&-e^{i c}&e^{i c} &-1&-1&e^{i c} &-e^{i c} &1\\
1&-e^{i a}&e^{ib} &-e^{if}&e^{if} &-e^{ib} &e^{i a} &-1\\
1&-e^{ia} &-e^{ib} &e^{if} &- e^{if}&e^{ib}& e^{ia} &-1\\
1&-1&-1&1&1&-1&-1&1
\end{array}\right]\label{D15}
\end{eqnarray}
and its polynomial is
\begin{eqnarray}
 P_{D_{8L}^{(5)}}(\lambda) &=&(1-\lambda^2)^2\left(4\,\lambda^4+ \sqrt{2}\,(e^{i(b+d-f)}-2\,e^{ia}+e^{ib})\lambda^3- 2(e^{i(a+b)}+e^{ i(c+f)})(e^{ i(d-f)}+1)\lambda^2\right.\nonumber\\
&&\left. +\sqrt{2}\,e^{i(c+f)}( e^{ i(a+d)}-2\,e^{ i(b+d)}+e^{i(a+f)})\lambda+4e^{ i(a+b+c+d)}\right)\label{8L}\end{eqnarray}
which is similar to that of the $D_{8F}^{(5)}$ matrix, see (\ref{8F}).

Looking carefully at the above polynomials we conclude that the $D_{8A}^{(5)}$, $D_{8D}^{(5)}$ and  $D_{8G}^{(5)}$   matrices generate a single class whose representative polynomial can be taken (\ref{8G!}) since the polynomials   (\ref{8A}), (\ref{8A!}), (\ref{8D}), (\ref{8G}) and (\ref{8G!}) depend on the same set of three distinct parameters, $a$,  $-a+b+c+d+f$ and $b+c+d+f$. The $D_{8H}^{(5)}$ and  $D_{8I}^{(5)}$ matrices, and respectively  $D_{8J}^{(5)}$ and  $D_{8K}^{(5)}$, define each one a new class, while  $D_{8B}^{(5)}$, $D_{8C}^{(5)}$, $D_{8E}^{(5)}$, $D_{8F}^{(5)}$ and $D_{8L}^{(5)}$ define five classes. 

Thus our proposal for equivalence of complex Hadamard matrices is the characteristic polynomial, which includes all the possible forms it could take by using  simple re-parametrization, and/or row and column permutations. The standard form of these eight classes is given in the next Section.

The spectra of matrices that depend on  four  phases are given by the following polynomials
\begin{eqnarray}
P_{D_{8A}^{(4)}}(\lambda)&=&(1-\lambda^2)^2\left(2\lambda^4-\sqrt{2}e^{i a}\lambda^3-\sqrt{2}e^{i( a+c+d)}\lambda+2e^{i(2 a+c+d)}\right)\\
P_{D_{8B}^{(4)}}(\lambda)&=&(1-\lambda^2)^2\left(2\lambda^4-\sqrt{2}e^{i(b- a)}\lambda^3-\sqrt{2}e^{i( a+c+d)}\lambda+2e^{i( b+c+d)}\right)\\
P_{D_{8C}^{(4)}}(\lambda)&=&(1-\lambda^2)^2\left(2\lambda^4-\sqrt{2}e^{i a}\lambda^3+ i\sqrt{2}e^{i( b+2d)}\lambda-2\,i\,e^{i( a+b+2d)}\right)\\
P_{D_{8D}^{(4)}}(\lambda)&=&(1-\lambda^2)^2\left(2\lambda^4-\sqrt{2}e^{i a}\lambda^3+ i\sqrt{2}e^{i( b+2c+2d-2a)}\lambda-2\,i\,e^{i( -a+b+2c+2d)}\right)\end{eqnarray}
which are similar to the preceding ones and show that the four matrices are not equivalent. 

 The matrix ${D_{8A}^{(3)}}$ leads to
\begin{eqnarray}
P_{D_{8A}^{(3)}}(\lambda)=(1-\lambda^2)^2\left(\lambda^4-\frac{1-i}{2\sqrt{2}}e^{i a}\lambda^3+\left(\frac{1-i}{2}e^{i(b+c)}-\frac{1+i}{2}e^{2i a}\right)\lambda^2
+\frac{1+i}{2\sqrt{2}}e^{i( a+b+c)}\lambda-e^{i(2 a+b+c)}\right)\end{eqnarray}

\section{Standard Form}

 For writing the complex Hadamard matrices in their standard form, \cite{TZ},  we need the real Hadamard matrices obtained by taking all the parameters zero in $D_{8*}^{(5)}$ , and the corresponding phase matrix  $R_{8*}^{(5)}$. 
There are four real Hadamard matrices
\begin{eqnarray}\begin{array}{cc}
h_1=\left[\begin{array}{crrrrrrr}
1&1&1&1&1&1&1&1\\
1&1&1&1&-1&-1&-1&-1\\
1&1&-1&-1&1&1&-1&-1\\
1&1&-1&-1&-1&-1&1&1\\
1&-1&1&-1&-1&1&-1&1\\
1&-1&-1&1&-1&1&1&-1\\
1&-1&1&-1&1&-1&1&-1\\
1&-1&-1&1&1&-1&-1&1
\end{array}\right], &
h_2=\left[\begin{array}{rrrrrrrr}
1&1&1&1&1&1&1&1\\
1&1&-1 &1 &-1  &-1 &1&-1\\
1&1 &1&-1 &1&-1&-1 &-1\\
1&1&-1&-1&-1&1&-1&1\\
1&-1&1&-1&-1&-1&1&1\\
1&-1&-1&-1&1&1&1&-1\\
1&-1&1&1&-1&1&-1&-1\\
1&-1&-1&1&1&-1&-1&1\\
\end{array}\right]\end{array}\label{hh1}
\end{eqnarray}

\begin{eqnarray}\begin{array}{cc}
h_3=\left[\begin{array}{rrrrrrrr}
1&1&1&1&1&1&1&1\\
1&1&1 &1 &-1  &-1 &-1&-1\\
1&1 &-1&-1 &1&-1&1 &-1\\
1&1&-1&-1&-1&1&-1&1\\
1&-1&1&-1&-1&-1&1&1\\
1&-1&1&-1&1&1&-1&-1\\
1&-1&-1&1&-1&1&1&-1\\
1&-1&-1&1&1&-1&-1&1\\
\end{array}\right], &
h_4=\left[\begin{array}{rrrrrrrr}
1&1&1&1&1&1&1&1\\
1&1&1 &1 &-1  &-1 &-1&-1\\
1&1 &-1&-1 &1&1&-1 &-1\\
1&1&-1&-1&-1&-1&1&1\\
1&-1&1&-1&-1&1&-1&1\\
1&-1&1&-1&1&-1&1&-1\\
1&-1&-1&1&-1&1&1&-1\\
1&-1&-1&1&1&-1&-1&1\\
\end{array}\right]\end{array}\label{hh2}
\end{eqnarray}
The corresponding $R_{8*}^{(5)}$ matrices have the form
\begin{eqnarray}
R_{8A}^{(5)}(a,b,c,d,f)=\left[\begin{array}{crrccrrr}
\bullet&\bullet&\bullet&\bullet&\bullet&\bullet&\bullet&\bullet\\
\bullet&a&f&d&d&f&a&\bullet\\
\bullet&b&f&b+d-a&b+d-a&f&b&\bullet\\
\bullet&c&c&\bullet&\bullet&c&c&\bullet\\
\bullet&c&c&\bullet&\bullet&c&c&\bullet\\
\bullet&c&f&b+d-a&b+d-a&f&b&\bullet\\
\bullet&a&f&d&d&f&a&\bullet\\
\bullet&\bullet&\bullet&\bullet&\bullet&\bullet&\bullet&\bullet
\end{array}\right]\label{r1}\end{eqnarray}
\begin{eqnarray}
R_{8B}^{(5)}(a,b,c,d,f)=\left[\begin{array}{crccccrr}
\bullet&\bullet&\bullet&\bullet&\bullet&\bullet&\bullet&\bullet\\
\bullet&a&a&d&d&a&a&\bullet\\
\bullet&b&b-c+f&d&d&b-c+f&b&\bullet\\
\bullet&c&f&\bullet&\bullet&f&c&\bullet\\
\bullet&c&f&\bullet&\bullet&f&c&\bullet\\
\bullet&b&b-c+f &d&d&b-c+f & b&\bullet\\
\bullet&a&a&d&d&a&a&\bullet\\
\bullet&\bullet&\bullet&\bullet&\bullet&\bullet&\bullet&\bullet
\end{array}\right]\label{r2} \end{eqnarray}

\begin{eqnarray}
R_{8C}^{(5)}(a,b,c,d,f)=\left[\begin{array}{crccccrr}
\bullet&\bullet&\bullet&\bullet&\bullet&\bullet&\bullet&\bullet\\
\bullet&a&a-c+f&d&d&a-c+f&a&\bullet\\
\bullet&b&b&d&d&b&b&\bullet\\
\bullet&c&f&\bullet&\bullet&f&c&\bullet\\
\bullet&c&f&\bullet&\bullet&f&c&\bullet\\
\bullet&b&b&d&d&b&b&\bullet\\
\bullet&a&a-c+f&d&d&a-c+f&a&\bullet\\
\bullet&\bullet&\bullet&\bullet&\bullet&\bullet&\bullet&\bullet
\end{array}\right]\label{r4}\end{eqnarray}

\begin{eqnarray}
R_{8E}^{(5)}(a,b,c,d,f)=\left[\begin{array}{crccccrr}
\bullet&\bullet&\bullet&\bullet&\bullet&\bullet&\bullet&\bullet\\
\bullet&a&f&a-b+d&a-b+d&f&a&\bullet\\
\bullet&b&f&d&d&f&b&\bullet\\
\bullet&c&c&\bullet&\bullet&c&c&\bullet\\
\bullet&c&c&\bullet&\bullet&c&c&\bullet\\
\bullet&b&f&d&d&f&b&\bullet\\
\bullet&a&f&a-b+d&a-b+d&f&a&\bullet\\
\bullet&\bullet&\bullet&\bullet&\bullet&\bullet&\bullet&\bullet
\end{array}\right]\label{r5}\end{eqnarray}

\begin{eqnarray}
R_{8F}^{(5)}(a,b,c,d,f)=\left[\begin{array}{crccccrr}
\bullet&\bullet&\bullet&\bullet&\bullet&\bullet&\bullet&\bullet\\
\bullet&a&f&d&d&f&a&\bullet\\
\bullet&a&f&d&d&f&a&\bullet\\
\bullet&c&c&\bullet&\bullet&c&c&\bullet\\
\bullet&c&c&\bullet&\bullet&c&c&\bullet\\
\bullet&a&b-d+f &b&b&b-d+f &a&\bullet\\
\bullet&a&b-d+f &b&b&b-d+f &a&\bullet\\
\bullet&\bullet&\bullet&\bullet&\bullet&\bullet&\bullet&\bullet
\end{array}\right]\label{r6}\end{eqnarray}

\begin{eqnarray}
R_{8I}^{(5)}(a,b,c,d,f)=\left[\begin{array}{crccccrr}
\bullet&\bullet&\bullet&\bullet&\bullet&\bullet&\bullet&\bullet\\
\bullet&a&f&d&d&f&a&\bullet\\
\bullet&b&b+f-a&d&d&b+f-af&a&\bullet\\
\bullet&c&c&\bullet&\bullet&c&c&\bullet\\
\bullet&c&c&\bullet&\bullet&c&c&\bullet\\
\bullet&&b+f-a&d&d&b+f-a&a&\bullet\\
\bullet&a&f &d&d&f &a&\bullet\\
\bullet&\bullet&\bullet&\bullet&\bullet&\bullet&\bullet&\bullet
\end{array}\right]\label{r7}\end{eqnarray}

\begin{eqnarray}
R_{8K}^{(5)(a,b,c,d,f)}=\left[\begin{array}{crccccrr}
\bullet&\bullet&\bullet&\bullet&\bullet&\bullet&\bullet&\bullet\\
\bullet&a&d&f&f&d&a&\bullet\\
\bullet&b&b+d-a&f&&b+d-a&b&\bullet\\
\bullet&c&c&\bullet&\bullet&c &c&\bullet\\
\bullet&c&c &\bullet&\bullet&c &c&\bullet\\
\bullet&b&b+d-a&f&f&b+d-a&b&\bullet\\
\bullet&a&d&f&f&d&a&\bullet\\
\bullet&\bullet&\bullet&\bullet&\bullet&\bullet&\bullet&\bullet
\end{array}\right]\label{r8}\end{eqnarray}

\begin{eqnarray}
R_{8L}^{(5)(a,b,c,d,f)}=\left[\begin{array}{crccccrr}
\bullet&\bullet&\bullet&\bullet&\bullet&\bullet&\bullet&\bullet\\
\bullet&a&b+d-f&d&d&b+d-f&a&\bullet\\
\bullet&a&b+d-f&d&d&b+d-f&a&\bullet\\
\bullet&c&c&\bullet&\bullet&c &c&\bullet\\
\bullet&c&c &\bullet&\bullet&c &c&\bullet\\
\bullet&a&b&f&f&b&a&\bullet\\
\bullet&a&b&f&f&b&a&\bullet\\
\bullet&\bullet&\bullet&\bullet&\bullet&\bullet&\bullet&\bullet
\end{array}\right]\label{r9}\end{eqnarray}

where $ \bullet$ means zero. The standard form is
\begin{eqnarray}\begin{array}{cc}
D_{8A}^{(5)}(a,b,c,d,f)=h_1\circ {\rm EXP}\left(i\cdot R_{8A}^{(5)}(a,b,c,d,f)\right),& D_{8B}^{(5)}(a,b,c,d,f)=h_2\circ {\rm EXP}\left(i\cdot R_{8B}^{(5)}(a,b,c,d,f)\right)\end{array}
\label{Dd1}\end{eqnarray}

 \begin{eqnarray}\begin{array}{cc}
D_{8C}^{(5)}(a,b,c,d,f)=h_3\circ {\rm EXP}\left(i\cdot R_{8C}^{(5)}(a,b,c,d,f)\right),& D_{8E}^{(5)}(a,b,c,d,f)=h_1\circ {\rm EXP}\left(i\cdot R_{8E}^{(5)}(a,b,c,d,f)\right)\end{array}\label{Dd2}
\end{eqnarray}

 \begin{eqnarray}\begin{array}{cc}
D_{8F}^{(5)}(a,b,c,d,f)=h_1\circ {\rm EXP}\left(i\cdot R_{8F}^{(5)}(a,b,c,d,f)\right), & D_{8I}^{(5)}(a,b,c,d,f)=h_4\circ {\rm EXP}\left(i\cdot R_{8I}^{(5)}(a,b,c,d,f)\right)\end{array}\label{Dd3}
\end{eqnarray}

\begin{eqnarray}\begin{array}{cc}
D_{8K}^{(5)}(a,b,c,d,f)=h_4\circ {\rm EXP}\left(i\cdot R_{8K}^{(5)}(a,b,c,d,f)\right), & D_{8L}^{(5)}(a,b,c,d,f)=h_4\circ {\rm EXP}\left(i\cdot R_{8L}^{(5)}(a,b,c,d,f)\right)\end{array}
\label{Dd4}\end{eqnarray}

In relation (\ref{Dd1}) $\circ $ denotes the entry wise Hadamard product. It is well known that matrices under Hadamard product and usual addition generate a commutative algebra such that all the usual functions like the exponential (\ref{Dd1}) are well defined.

\section{Conclusion}

The use of the nonlinear doubling formula (\ref{con}) led to an entire new class of eight dimensional  complex Hadamard matrices which depend on five arbitrary phases. We did not make an exhaustive search for finding all such matrices, however  we found many matrices, and tried  to solve the problem of their equivalence by using the intrinsic properties of unitary matrices. By using the usual equivalence, (\ref{ha1}), their number grows very quickly. Only the eight matrices (\ref{Dd1})-(\ref{Dd4}), by using re-parametrization of the form $a\rightarrow a \pm \pi/2 $ for the free parameters entering a given matrix, provide $8\times 32=246$ non equivalent matrices, that is a big number. 

The use of orthogonality relations can solve a part of the equivalence problem, and an important result is that one matrix, its transpose and complex conjugate are equivalent. In this case comparing the fingerprints of two matrices is not an easy problem because the number of generated polygons if of the order $n(n-1)$ for each of the two matrices of order $n$. On the other hand the characteristic polynomial requires only one computation and, as the relations (\ref{8A})-(\ref{8L}) show,   the characteristic polynomial depends only on a few combinations of the same set of five parameters $(a,b,c,d,f)$, which are different for each class as defined before.
This new equivalence definition includes the orthogonality relations equivalence since it is easily seen that because of the properties of normal matrices,  the matrices, $H$, and  $H^t$, have the same characteristic polynomial,  and that of  $H^*$ is obtained by changing $\lambda\rightarrow 1/\lambda$ in the previous ones, which leads to the complex conjugate spectrum. In this way the number of non equivalent matrices is significantly lowered.

In principle the same equivalence could be applied to real Hadamard matrices. However a simple spectral computation of the $h_1,\,h_2,\,h_3,\,h_4$  matrices shows that only the matrices $h_1$ and $h_3$ are equivalent, and  $h_1$ is not equivalent to $h_2$ and $h_4$, nor  $h_2$ is  equivalent to  $h_4$. Their spectra are
\begin{eqnarray}
Sp(h_1)&=&Sp(h_3)=\left(-1^3,1^3,\frac{\sqrt{2}\pm i\sqrt{14}}{4}\right),\;\;Sp(h_4)=(-1^4,1^4)\nonumber\\
Sp(h_2)&=&\left(-1^2,1^2,\frac{1}{4}\left(\frac{1+\sqrt{17}}{\sqrt{2}}\pm i \sqrt{7-\sqrt{17}} \right),\frac{1}{4}\left(\frac{1-\sqrt{17}}{\sqrt{2}}\pm i \sqrt{7+\sqrt{17}} \right)\right)
\end{eqnarray}
where power means multiplicity. But taking into account that for real Hadamard matrices with dimension $d=2,\,4,8,\,12$   there is only one matrix under the usual equivalence, we do not suggest the use of the new equivalence because it will cause dramatic changes in the field. 

In contradistinction to the real case the parametrization of complex Hadamard matrices is still at its infancy stage, such that the use of the proposed equivalence does not imply major changes.

One important problem which deserves to be solved is the finding of all the possible classes of eight-dimensional complex Hadamard matrices that depend on five arbitrary phases. We used an ad-hoc method to find the above results, and presumably there are more. Closed related ones would be the description of the corresponding orbits, are they related, or not, to the form of characteristic polynomials, and solving the mystery why the use of the $H_4(a)$ matrix leads to matrices which depend on four arbitrary phases, similar to (\ref{D12}) and  (\ref{D13}), that cannot be brought to forms which do not explicitly depend on the imaginary unit.

\begin{acknowledgments}  We acknowledge partial support from  Program Idei, contract no 464/2009, and  ANCS contract no 15EU/06.04.2009\end{acknowledgments}


\end{document}